\documentclass{article}
\usepackage{bm,latexsym,amsmath,amssymb,amsfonts,fancyhdr,color,graphicx,multirow,slashed,cite,bbold,cancel,booktabs}
\usepackage[a4paper,bottom=3cm,top=2.5cm,head=0mm,width=17cm,dvipdfm]{geometry}
\usepackage[usenames,dvipsnames,svgnames,table]{xcolor}
\usepackage[colorlinks=true,
            linkcolor=blue,
            urlcolor=blue,
            citecolor=green,          
            bookmarks=true,
            bookmarksnumbered=true,
            breaklinks=true,
            pdfstartview=FitBH
            ]{hyperref}
\usepackage[dotinlabels]{titletoc}
\usepackage{titlesec}
\usepackage{authblk,ulem}

\numberwithin{equation}{section}
\allowdisplaybreaks[4]

\titlelabel{\thetitle.\quad \hspace{-0.8em}}
\titlecontents{section}
              [1.5em]
              {\vspace{4mm} \large \bf}
              {\contentslabel{1em}}
              {\hspace*{-1em}}
              {\titlerule*[.5pc]{.}\contentspage}
\titlecontents{subsection}
              [3.5em]
              {\vspace{2mm}}
              {\contentslabel{1.8em}}
              {\hspace*{.3em}}
              {\titlerule*[.5pc]{.}\contentspage}
\titlecontents{subsubsection}
              [5.5em]
              {\vspace{2mm}}
              {\contentslabel{2.5em}}
              {\hspace*{.3em}}
              {\titlerule*[.5pc]{.}\contentspage}

\newcommand{\titledef}{Identifying Neutrino Mass Ordering with Cosmic Gravitational Focusing} % Insert Title here!!!
\hypersetup{ pdfauthor = {Shao-Feng Ge},
	     pdftitle = {\titledef}, % Insert title here!!!
	     pdfsubject = {}, % Insert subject here!!!
             pdfkeywords = {}, % Insert keywords here!!!
	     pdfcreator = {LaTeX with hyperref package},
	     pdfproducer = {dvips + ps2pdf} }

\definecolor{gesfblack}{rgb}{0,0,0}

\definecolor{gesfblue}{rgb}{0.08,0.42,0.76}

\definecolor{gesfgreen}{rgb}{0,1,0}

\definecolor{gesfgrey}{rgb}{0.5,0.5,0.5}

\definecolor{gesflanse}{rgb}{0.00,0.50,0.50}

\definecolor{gesfpurple}{rgb}{0.47,0.19,0.42}

\definecolor{gesfred}{rgb}{1,0,0}
\newcommand{\gred}[1]{{\color{gesfred} #1}}
\definecolor{gesfwhite}{rgb}{1,1,1}

\definecolor{gesfyellow}{rgb}{0.7,0.4,0.3}

\newcommand{\gsec}[1]{{\hypersetup{linkcolor=red}Sec.\,\ref{#1}\hypersetup{linkcolor=blue}}}

\newcommand{\geqn}[1]{\hypersetup{linkcolor=blue}Eq.\,(\ref{#1})\hypersetup{linkcolor=blue}}
\newcommand{\gfig}[1]{{\hypersetup{linkcolor=violet}Fig.\,\ref{#1}\hypersetup{linkcolor=blue}}}

\definecolor{Orange}{cmyk}{0,0.61,0.87,0}
\definecolor{JungleGreen}{cmyk}{0.99,0,0.52,0}
\definecolor{OliveGreen}{cmyk}{0.64,0,0.95,0.40}
\definecolor{Brown}{cmyk}{0,0.81,1,0.60}
\definecolor{RoyalBlue}{cmyk}{0.71,0.53,0,0.12}
\definecolor{Gray}{cmyk}{0,0,0,0.40}
\definecolor{LightPink}{cmyk}{0.0,0.25,0,0}
\definecolor{LLightPink}{cmyk}{0.0,0.10,0,0}
\definecolor{LightBlue}{cmyk}{0.25,0,0,0}
\definecolor{LightGray}{cmyk}{0,0,0,0.2}

\definecolor{Plum}{rgb}{0.56, 0.27, 0.52}

\usepackage{float}

\graphicspath{{Figures/}}

\begin{document}
\fontsize{12pt}{14pt}\selectfont

\title{%\begin{flushright}
       %\mbox{\normalsize IPMU18-xxxx}
       %\end{flushright}
			 %\vskip 20pt
       \textbf{\fontsize{21pt}{22pt}\selectfont \titledef}} % Insert title here!!!
\author[1,2]{{\large Shao-Feng Ge} \footnote{\href{mailto:gesf@sjtu.edu.cn}{gesf@sjtu.edu.cn}}}
\affil[1]{State Key Laboratory of Dark Matter Physics, Tsung-Dao Lee Institute \& School of Physics and Astronomy, Shanghai Jiao Tong University, Shanghai 200240, China}
\affil[2]{Key Laboratory for Particle Astrophysics and Cosmology (MOE) \& Shanghai Key Laboratory for Particle Physics and Cosmology, Shanghai Jiao Tong University, Shanghai 200240, China}
\author[1,2]{{\large Liang Tan} \footnote{\href{mailto:tanliang@sjtu.edu.cn}{tanliang@sjtu.edu.cn}}}

\date{\today}

\maketitle

\begin{abstract}
\fontsize{12pt}{14pt}\selectfont

The cosmic gravitational focusing (CGF) of relic neutrinos
can provide an independent measurement of the absolute neutrino masses $m_i$
with fourth-power dependence ($m^4_i$).
We demonstrate in this paper for the first time
how this can help identifying the neutrino mass ordering,
using the fact that total mass falling below the inverted ordering threshold
allows the discrimination of the inverted ordering.
Upon incorporating the projected CGF sensitivity at DESI,
the preference for the normal ordering with
a prior $\sum m_i > 0.059\,{\rm eV}$ would increase
from the original 89.9\% of the
existing matter clustering method with the
DESI analysis to 98.2\%
while the inverted ordering
is further disfavored from 10.1\% to 1.8\%. 
We also show how this can affect the prospects
of the neutrinoless double beta decay and single
beta decay measurements.

\end{abstract}

\newpage

\section{Introduction}
\label{sec:intro}

The neutrino oscillation 
\cite{Super-Kamiokande:1998kpq,SNO:2002tuh}
is the first experimentally verified 
new physics beyond the Standard Model of particle physics \cite{Mohapatra:2005wg,Gonzalez-Garcia:2007dlo,Giganti:2017fhf}.
The neutrino oscillation experiments are sensitive to the neutrino 
mass squared differences
$\Delta m^2_{ij} \equiv m^2_i - m^2_j$. 
The current global fit gives 
the atmospheric mass squared difference
$ |\Delta m_a^2 \equiv \Delta m_{31}^2 | \approx 2.517 \times 10^{-3}\,{\rm eV}^2 $ 
and the solar mass squared difference
$ \Delta m_s^2 \equiv \Delta m_{21}^2 \approx 7.42 \times 10^{-5}\,{\rm eV}^2$ \cite{Esteban:2020cvm}.
However, the absolute neutrino mass and mass ordering (the sign of $\Delta m_{31}^2$) are unknown,
leaving the two cases of normal ordering (NO) $ \Delta m^2_{31} > 0$
and inverted ordering (IO) $\Delta m^2_{31}<0$ to be determined.
The current neutrino oscillation global fit
has excluded IO by 2$\sim$2.7$\sigma$ 
\cite{deSalas:2020pgw,Esteban:2020cvm,Capozzi:2021fjo}.
The JUNO experiment aims to exclude the IO by nearly 3\,$\sigma$ with six years of running \cite{JUNO:2015zny,JUNO:2015sjr}
and can further improve to 4\,$\sigma$ if including
atmospheric neutrino oscillation \cite{Zhang:2021adu}.

The neutrino mass ordering (NMO) is important for
the single and neutrinoless double beta decays.
The single beta decay experiments can measure the absolute neutrino mass through 
its spectrum around the endpoint 
with a dependence on the
combination $m_\beta \equiv \sum | U_{e i} |^2 m_i$ where $U_{e i}$ are the (Pontecorvo–Maki–Nakagawa–Sakata) PMNS
matrix elements \cite{ParticleDataGroup:2022pth}.
The current KATRIN bound is $m_{\beta} < 0.45\,{\rm eV}$ \cite{KATRIN:2021uub,Katrin:2024tvg}
with the near-future target of $m_{\beta} < 0.2\,{\rm eV}$ \cite{KATRIN:2005fny}.
The future Project 8 is expected to touch the bottom of the NO prediction
around $0.04\,{\rm eV}$ \cite{Project8:2014ivu}.
If neutrinos are the Majorana type, the neutrinoless double beta decay
($0\nu\beta \beta$) may occur
whose reaction rate
depends on the effective mass
$m_{\beta \beta} \equiv \sum U_{e i}^2 m_i$. 
The current experiments KamLAND-Zen \cite{KamLAND-Zen:2022tow} and
GERDA \cite{GERDA:2020xhi} constrain $m_{\beta \beta}$
to be less than $(36 \sim 156 )\,{\rm meV}$ and $(79 \sim 180 )\,{\rm meV}$, respectively. 
A future proposed experiment aims to reach 
$m_{\beta \beta} = ( 18.4 \pm 1.3 )\,{\rm meV}$ \cite{Agostini:2021kba}.

Comparing with these underground experiments,
the existing matter clustering methods
\cite{Lesgourgues:2006nd,TopicalConvenersKNAbazajianJECarlstromATLee:2013bxd,Dvorkin:2019jgs,DiValentino:2024xsv},
including both cosmic microwave background (CMB) and large scale structure (LSS) observations,
can provide a more sensitive measurement of the absolute neutrino mass and mass ordering 
\cite{Planck:2018vyg,DESI:2024mwx}.
The current cosmology data favor NO over IO \cite{Hannestad:2016fog,Vagnozzi:2017ovm,Tanseri:2022zfe}, given a physical prior $\sum m_i > 0.059\,{\rm eV}$.
There are also tries of extending the analysis to the
unphysical region $\sum m_i < 0$ of negative neutrino masses
\cite{Craig:2024tky,Green:2024xbb,Herold:2024enb}
\footnote{Note that the quantity of Standard Model matter that enters
the general relativity (GR) is the energy momentum tensor
$T^{\mu \nu}$. Only in the nonrelativistic limit that
the neutrino energy ($\sim T^{00}$) reduces to the
absolute value of the neutrino mass. Although the fermion
mass can indeed be negative by chiral rephasing, the
one probed by GR and hence cosmology is its absolute
value. For example, $E_\nu \approx |m_\nu| + m^2_\nu / 2 p_\nu$. The negative mass scenario is actually a negative
energy one.}.
More advanced analysis with late-time background probes
can be found in \cite{Wang:2024hen,Jiang:2024viw}.
The over-stringent cosmic neutrino mass constraint
might arise from the CMB lensing anomaly
\cite{RoyChoudhury:2019hls,Motloch:2019gux,DiValentino:2021imh,Allali:2024aiv,Naredo-Tuero:2024sgf}.
Or it can be alleviated by dynamical dark energy \cite{Yadav:2024duq,Du:2024pai,Elbers:2024sha,Reboucas:2024smm,Shao:2024mag},
dark matter (DM) long-range force \cite{Craig:2024tky}, and
neutrino dark matter interaction \cite{Sen:2024pgb}.
In addition, it is widely explored to weaken the cosmological neutrino mass constraint in other new physics scenarios,
such as primordial non-Gaussian \cite{Forconi:2023akg},
modified gravity \cite{Bellomo:2016xhl},
time varying neutrino mass \cite{Lorenz:2018fzb},
neutrino non-thermal distribution \cite{Oldengott:2019lke,Alvey:2021sji},
neutrino decay \cite{Beacom:2004yd,Farzan:2015pca,Chacko:2019nej,Chacko:2020hmh,Escudero:2020ped},
neutrino annihilation \cite{Escudero:2022gez},
and neutrino self-interaction \cite{Esteban:2021ozz}.

Since the neutrino effect on the matter clustering 
arises from their contributions to the
cosmic energy density that are dominated by the neutrino masses
after entering the nonrelativistic regime, what
these existing matter clustering methods
can probe is the total sum of the
neutrino masses $\sum m_i$.
For comparison, the CGF effect \cite{Ge:2023nnh} that uses the
relative velocity and gravitational attraction \cite{Zhu:2013tma,Zhu:2014qma,Inman:2015pfa,Okoli:2016vmd,Zhu:2019kzb,Nascimento:2023ezc} between
the cosmic neutrino fluid (C$\nu$F) and DM halos,
is sensitive to the fourth power of the neutrino masses $m^4_i$ \cite{Okoli:2016vmd,Ge:2023nnh}.
This provides a complementary approach in cosmology to measure
the neutrino masses \cite{Ge:2023nnh}.

Although the projected CGF sensitivity on
the absolute neutrino mass measurement has been detailed in \cite{Ge:2023nnh},
its role of distinguishing the two NMOs has not been elaborated yet.
\gsec{sec:nu_ordering} evaluates the relative
probabilities of the two NMOs with
both the existing matter clustering methods and its CGF counterpart,
based on the fact that the total mass falling below the IO threshold 
would disfavor IO.
\gsec{sec:CGF_TE} further studies its effect
on the terrestrial experiments of single and neutrinoless
double beta decays. Our conclusions can be found
in \gsec{sec:conclusion}.

\section{Cosmological Measurements of the Neutrino Mass Ordering}
\label{sec:nu_ordering}

The existing cosmological measurements of neutrino masses
based on the matter clustering take the observations of the
CMB and LSS. In both phenomena,
the neutrino effect on the existing matter clustering methods is
mainly contributed by its energy density which is
effectively the neutrino mass in the nonrelativistic
regime.
As detailed in \cite{Ge:2023nnh}, the CGF effect
is intrinsically different from
the existing matter clustering method. Due to the gravitational
focusing sourced by the DM halos, the C$\nu$F
develops a density dipole. With larger mass,
the neutrino deflection in the gravitational potential
increases much faster than linear response. For
nonrelativistic neutrinos, the density dipole receives
a fourth-power dependence $m^4_i$ on the neutrino masses
$m_i$ which allows improvement on the neutrino mass
measurement from CGF. Since the neutrino mass measurement
with the existing matter clustering methods prefers the
mass region below the IO threshold and hence disfavors the IO
\cite{Hannestad:2016fog, Ge:2019ldu}, the CGF mass measurement
can also be used to identify the NMO.
As CGF can have a comparable sensitivity for the neutrino
mass measurement, one may expect the same happens for
the NMO determination. The CGF effect is truly a complementary
method than the existing matter clustering methods, for
not just the neutrino mass measurement but also the NMO.

\gsec{sec:DESI_nu} briefly discusses the NMO measurement
from the newly released DESI baryonic acoustic oscillation
(BAO) data \cite{DESI:2024mwx} 
combined with the existing CMB anisotropies from Planck \cite{Planck:2019nip}
and CMB lensing data from Planck \cite{Carron:2022eyg} and ACT \cite{ACT:2023kun,ACT:2023dou,ACT:2023ubw}.
We abbreviate this data combination as the "{\it DESI analysis}" 
in the remaining part of this paper.
For comparison, the projected sensitivity with only CGF
is elaborated in \gsec{sec:ordering_CGF}
while its combination with
the existing matter clustering methods can be found in
\gsec{sec:CGF_combine}.

\subsection{Matter clustering methods}
\label{sec:DESI_nu}

Currently, cosmological observation
is the most stringent way to constrain the absolute neutrino mass. 
Combining the recent first year DESI data release of its BAO
observation and the previous CMB data,
the upper limit on the neutrino mass sum can reach
$\sum m_i < 0.072\,{\rm eV}$ 
at 95\%\,C.L. for a positive mass sum
$\sum m_i > 0$ prior \cite{DESI:2024mwx}.
If a more physical prior is considered,
$\sum m_i > 0.059\,(0.101)\,{\rm eV}$ for NO (IO),
the 95\%\,C.L. is $\sum m_i < 0.113\,(0.145)\,{\rm eV} $.
With full data collection, 
DESI aims to achieve a sensitivity of $\sigma_{\Sigma} = 30\,{\rm meV}$ 
\cite{Ravoux:2024gtk}. 
We would take the existing measurements as input for
the following analysis in this section.

\begin{figure}[t]
\centering
\includegraphics[width=0.49\textwidth]{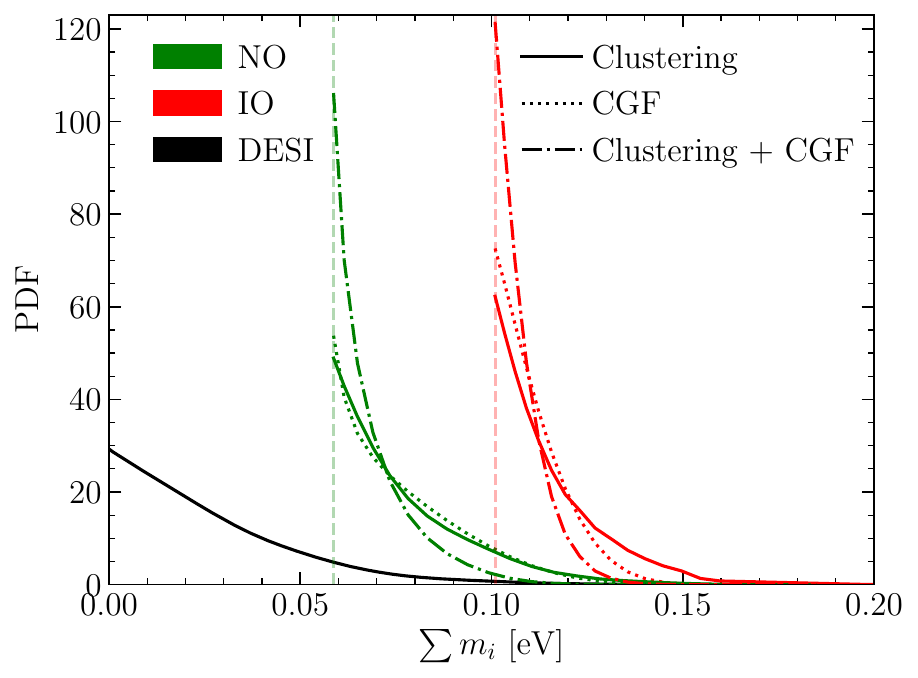}
\includegraphics[width=0.49\textwidth]{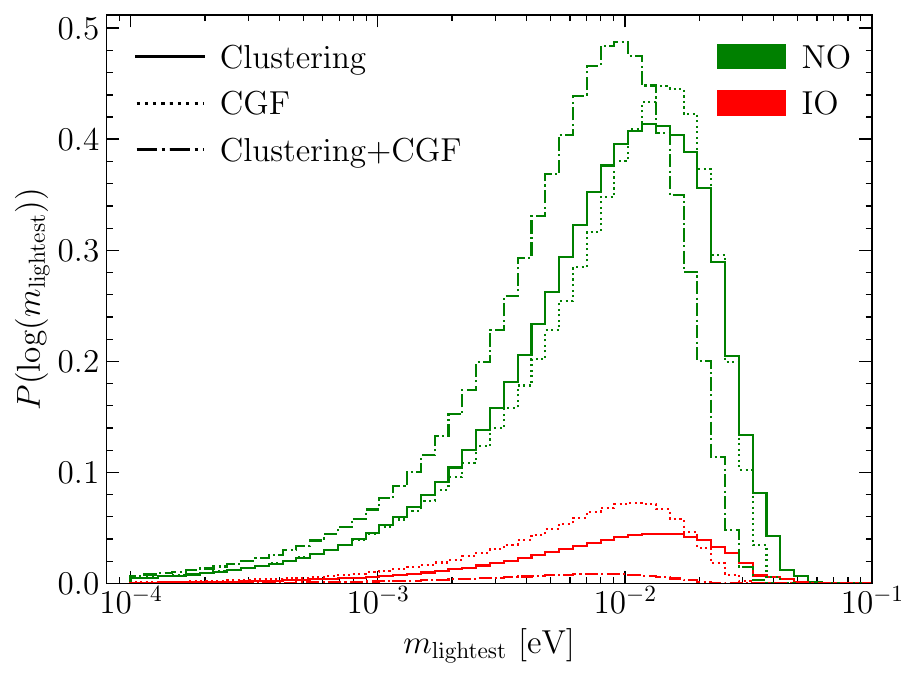}
\caption{
{\bf Left}:  
The probability density function (PDF) from the DESI analysis is shown as black solid line \cite{DESI:2024mwx}. 
The results by imposing NO and IO are
shown with green solid and red solid lines, respectively. 
For comparison, the PDFs of CGF and its combination with the clustering method
are also shown as dotted and dash-dotted lines, respectively.
The vertical dashed lines at 0.059\,eV and 0.101\,eV represent the minimum mass sum for NO (green) and IO (red). 
{\bf Right}:  
The distribution of the lightest neutrino mass sampled for NO (green) and IO (red) from the neutrino mass sum
PDF from the clustering method (solid), 
CGF (dotted), and their combination (solid-dotted).
}
\label{fig:pdf_DESI_DR1}
\end{figure}

The probability density function (PDF) for the neutrino mass sum $P (\Sigma m_i)$,
extracted from the DESI analysis \cite{DESI:2024mwx},
is plotted as the black line in \gfig{fig:pdf_DESI_DR1}.
To make it directly usable, we have normalized this PDF to 1,
$ \int_0^{\infty} P (\sum m_i) d (\sum m_i) =1  $,
in the physical range $\sum m_i > 0$.
The original PDF variable, namely, the mass sum $\sum m_i$,
can be replaced by
the lightest neutrino mass $m_1$ ($m_3$) for NO (IO),
respectively,
\begin{subequations}
\begin{align}
  \Sigma_{\rm NO}
& \equiv
  m_1
+ \sqrt{ m_1^2 + m_s^2}
+ \sqrt{ m_1^2 + m_a^2}, 
\\
  \Sigma_{\rm IO}
& \equiv
  m_3
+ \sqrt{ m_3^2 + m_a^2}
+ \sqrt{ m_3^2 + m_a^2 + m_s^2}.
\end{align} 
\end{subequations}
Correspondingly, the neutrino mass PDF becomes,
\begin{align} 
  P_{\rm NO} (m_1)
\propto
  P\left( \Sigma_{\rm NO} \right) 
  \frac{d \Sigma_{\rm NO}}{d m_1},
\quad {\rm and} \quad 
  P_{\rm IO} (m_3)
\propto
  P\left( \Sigma_{\rm IO} \right)
  \frac{d \Sigma_{\rm IO}}{d m_3},
\end{align} 
according to the Jacobian rule. After variable
transformation, the original physical region
$\sum m_i > 0$ can no longer be fully filled. It is then
necessary to renormalize the transformed PDF's to 1,
$ \int_0^\infty d m_1 P_{\rm NO} (m_1) = 1 $
and $ \int_0^\infty d m_3 P_{\rm IO} (m_3) = 1 $.
The normalized
$P (\Sigma_{\rm NO})$ and $P (\Sigma_{\rm IO})$
are shown as
green and red solid lines in the left panel of
\gfig{fig:pdf_DESI_DR1}, respectively,
for direct comparison with the original PDF (black solid).

With two options or values, the NMO
should be described by a discrete variable.
The probability of a particular mass ordering should be represented by the
corresponding relative chance 
with a flat prior assumption on the lightest neutrino mass 
\cite{Hannestad:2016fog, Ge:2019ldu},
\begin{align} 
  P_{\rm NO / IO}
\equiv
  \frac{  \int d m_{1/3} ( d \Sigma_{\rm NO/IO} / d m_{1/3} ) P (\Sigma_{\rm NO/IO})  } 
       { \int d m_1 ( d \Sigma_{\rm NO} / d m_1 ) P (\Sigma_{\rm NO})  +  \int d m_3 ( d \Sigma_{\rm IO} / d m_3 ) P (\Sigma_{\rm IO}) }.
\label{eq:p_IO_NO}
\end{align} 
With this definition, the two probabilities sum up to 100\%,
$P_{\rm NO} + P_{\rm IO} = 1$.
Combining the original PDF $P (\sum m_\nu)$
(the black line in \gfig{fig:pdf_DESI_DR1})
from the DESI analysis \cite{DESI:2024mwx}
with the measured neutrino mass squared differences
from the terrestrial oscillation experiments \cite{KamLAND:2013rgu,T2K:2020nqo,ParticleDataGroup:2022pth}
gives,
\begin{align} 
  P_{{\rm NO, DESI}}
& \approx 89.9\%
,\quad {\rm and} \quad
  P_{{\rm IO, DESI}}  
\approx 10.1\%.
\label{eq:PNOPIO}
\end{align}
This indicates a 1.64\,$\sigma$ preference of NO \cite{DESI:2024mwx},
Here we assume that the neutrino oscillation experiments
have already fixed the mass squared differences and
neglect their uncertainties for simplicity.
With the PDFs from clustering (solid) in \gfig{fig:pdf_DESI_DR1}
and the relative probabilities in \geqn{eq:PNOPIO},
the lightest neutrino mass distributions
$P_{\rm NO} (\log(m_1))$ and $P_{\rm IO} (\log(m_3))$ for NO (green) and IO (red) are sampled in the right panel of \gfig{fig:pdf_DESI_DR1},
The maximum probability emerges near 
$m_{\rm lightest} \approx 0.01\,{\rm eV}$ for both NO and IO.

\subsection{Cosmic gravitational focusing}
\label{sec:ordering_CGF}

In addition to the existing matter clustering methods discussed in \gsec{sec:DESI_nu},
CGF can give another 
independent cosmological way of measuring the neutrino mass.
The CGF effect arises from
the gravitational attraction and relative velocity between the
C$\nu$F and DM halo.
The combined effect
focuses the neutrinos downstream and induces a dipole density
distribution
to be traced by the galaxy cross-correlation functions \cite{Zhu:2013tma,Ge:2023nnh}.
Actually, the relative velocity arises from the nonlinear evolution of gravity
and it manifests in the three-point (3pt) correlation functions \cite{Zhu:2019kzb}.
So the CGF effect can be evaluated as
the squeezed limit of the 3pt correlation functions \cite{Ge:2023nnh}.
In this section, we further show explicitly that
CGF can significantly improve the measurement of the NMO.

The CGF induces the density fluctuation along the velocity of
neutrino fluids. Moving to the Fourier space, it manifests as \gred{an}
imaginary part in the neutrino density fluctuation,
$\tilde \delta_m  =( 1 + i \tilde \phi_\nu ) \tilde \delta_{m 0}$,
where $\tilde \delta_{m 0}$ is the matter overdensity
in the absence of CGF.
Its imaginary part induced by CGF is parameterized as
$i \tilde \phi_\nu \tilde \delta_{m 0}$.
By solving the Boltzmann equation semi-analytically, the $\tilde \phi_\nu$ contains the
neutrino mass fourth-power dependence for nonrelativistic neutrinos 
\cite{Okoli:2016vmd,Ge:2023nnh},
\begin{align}
  \widetilde \phi_\nu 
\approx &
- \frac{2 G a^2}{|\bm k|^2}
  \frac{m_\nu^4 ({\bm v}_{\nu c} \cdot \hat {\bm k})}{e^{m_\nu |\bm v_{\nu c}\cdot \hat{\bm k}|/T_\nu} + 1},
\label{eq:phiNonRel}  
\end{align}
where $G$ is the Newton constant, $a$ is the scale factor, $m_\nu$ is the neutrino mass, and
$\bm v_{\nu c}$ is the neutrino relative velocity to DM \gred{halos}.

Since both neutrino and dark matter are invisible,
we use galaxies as tracer to extract the imaginary
part $\tilde \phi_\nu$ induced by CGF.
The galaxy number density fluctuation can be expressed
in terms of the matter overdensity $\tilde \delta_{m0}$
and the imaginary part carried by the cosmic neutrinos,
$ \tilde \delta_{g, \alpha} 
= 
  b_\alpha \tilde \delta_{m 0} 
+ i b_\nu \tilde \phi_\nu \tilde \delta_{m 0}$,
where $b_\alpha$ is the bias of the galaxy type $\alpha$
and the neutrino bias $b_\nu$ is close to 1 \cite{LoVerde:2014pxa}.
The signal is defined as the imaginary part of galaxy power spectrum,
$
  \mathcal S
\equiv
  {\rm Im} P_g
=
  {\rm Im}
  \langle  \tilde \delta_{g \alpha} \tilde \delta_{g \beta}  \rangle,
$
while the corresponding noise is defined as its variance 
$ \mathcal N  
\equiv  
\sqrt{
  \langle \mathcal S^2 \rangle - \langle \mathcal S \rangle^2 
}$.
We can get the signal-to-noise ratio (SNR) of CGF as \cite{Ge:2023nnh},
\begin{align}
  \left( \frac{\mathcal S}{\mathcal N} \right)^2 
=
  \sum_{z_i,\nu_j} 
  V_i
  \int \frac{d^3{\bm k}}{(2\pi)^3} 
  \frac{2\Delta b }{\rm Det[C]}
  \left \langle 
\left[ 
  \mu_{\bm k}^2  
  \frac{\dot{\widetilde \phi}_{\nu_j} } H  
  + (f \mu_{\bm k}^2 + 1 ) \widetilde \phi_{\nu_j}
\right]
  \widetilde \delta_{m0}^2
  \right \rangle^2,
\label{eq:LikelihoodDef}
\end{align}
where $V_i$ is the survey volume of the $i$-th redshift bin.
Note that the signal vanishes with $\Delta b$,
the bias difference between two different types of
galaxies. The normalization factor
${\rm Det} C$ is the determinant of the covariance matrix 
$C_{\alpha \beta} \equiv \langle \delta_{g, \alpha} \delta_{g, \beta} \rangle $.
In addition, the total SNR is obtained by
summing over the contribution
from the three neutrino masses, each labeled by the index $j$.

\begin{figure}[t]
\centering
\includegraphics[width=0.69\textwidth]{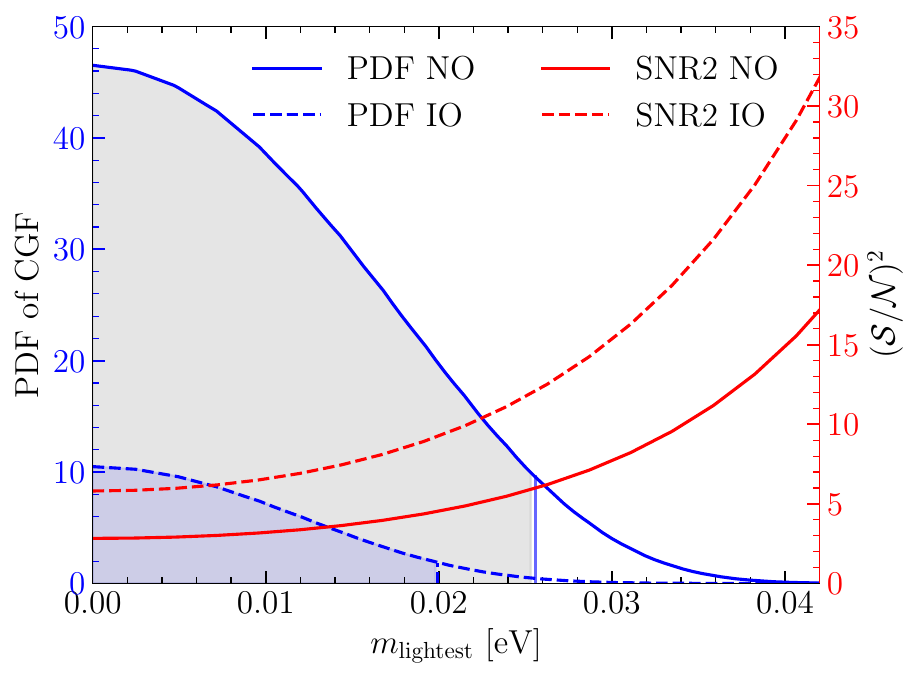}
\caption{
The CGF SNR squared of NO (solid) and IO (dashed) extracted from \cite{Ge:2023nnh}
are shown as red lines for the right axis.
For direct comparison with
the DESI analysis results, the PDFs of CGF
are also shown as blue lines
for NO (solid) and IO (dashed)
according to the left axis,
whose enclosed area with the corresponding
vertical lines indicate the 95\%\,C.L. regions.
}
\label{fig:PDF_CGF}
\end{figure}

We extract the SNR
squared $(\mathcal{S}/\mathcal{N})^2$ of CGF
from our previous studies \cite{Ge:2023nnh}, by using \geqn{eq:LikelihoodDef},
shown as red lines
for both NO (solid) and IO (dashed) in \gfig{fig:PDF_CGF}.
The corresponding PDF can be directly obtained from this SNR,
\begin{align} 
  P_{\rm CGF} (m_{\rm lightest}, {\rm NO/IO})
\equiv
 N_{\rm NO/IO}
 \exp \left[  - \frac 1 2 \left( \frac{\mathcal S}{\mathcal N} \right)^2   \right],
\end{align} 
where $m_{\rm lightest}$ stands for the $m_1$ ($m_3$) for NO (IO).
The normalization factors $N_{\rm NO/IO}$
are defined such that,
% $ \int d m_{\rm lightest} P_{\rm CGF} (m_{\rm lightest}, {\rm NO/IO}) = 1$.
$ \int d m_{\rm lightest} 
[ P_{\rm CGF} (m_{\rm lightest}, {\rm NO}) +  P_{\rm CGF} (m_{\rm lightest}, {\rm IO}) ] = 1$.
Here, we have assumed that the lightest neutrino mass is randomly distributed in the linear scale as done in \geqn{eq:p_IO_NO}.
The PDFs of CGF are shown as blue lines
in \gfig{fig:PDF_CGF} for both NO (solid) and IO (dashed).
% \gred{
% To make a comparison, we also plot the CGF PDF in terms of the mass sum
% variable in the right panel of \gfig{fig:PDF_CGF}.
% }

To make a comparison, the PDFs of CGF are also plotted as dotted lines
in the left panel of \gfig{fig:pdf_DESI_DR1} for
both NO (green) and IO (red),
after variable transformation from the lightest neutrino mass to 
the neutrino mass sum $\sum m_i$ and
adding the inverse Jacobian $ ( d \Sigma_{\rm NO/IO} / d m_{\rm lightest} )^{-1} $.
Note that the CGF PDF
decreases faster than the existing matter clustering one.
This behavior arises from the
fact that CGF has a fourth power
dependence on the neutrino masses \cite{Ge:2023nnh}
which increases faster with larger neutrino masses.
The PDFs of CGF give the 95\%\,C.L. limits
for the neutrino mass sum,
$\sum m_\nu < 0.109\,{\rm eV}$ and $\sum m_\nu < 0.129\,{\rm eV}$ for
NO (with prior $\sum m_\nu > 0.059\,{\rm eV}$) and IO
(with prior $\sum m_\nu > 0.101\,{\rm eV}$), respectively.
For IO, the projected CGF
sensitivity is slightly better than the DESI 
analysis result as discussed in the first paragraph
of \gsec{sec:DESI_nu}.

Using \geqn{eq:p_IO_NO} again, 
we can get the relative probabilities for
NO and IO with CGF only,
\begin{align} 
  P_{\rm NO, CGF}
& \approx 
  85.7\%
,\quad {\rm and} \quad
  P_{\rm IO, CGF}
\approx 
  14.3\%,
\label{eq:PNO_IO_CGF}
\end{align} 
which are proportional to the areas below the blue lines in \gfig{fig:PDF_CGF}.
These results are comparable with the current DESI analysis result in \geqn{eq:PNOPIO}.
There is a 1.46\,$\sigma$
preference of NO by using CGF alone.
 
With the PDF from the CGF measurements in \gfig{fig:PDF_CGF},
we sample the lightest neutrino
mass distribution with the
relative probabilities in \geqn{eq:PNO_IO_CGF}.
These sampled distributions are shown
as dotted lines in the right panel of \gfig{fig:pdf_DESI_DR1}
for both NO (green) and IO (red).
The maximum probability for the lightest neutrino mass
appears at slightly larger neutrino mass than the
$0.01\,{\rm eV}$ from the DESI analysis.

\subsection{Combination of clustering and cosmic gravitational focusing}
\label{sec:CGF_combine}

The existing matter clustering methods and CGF
are independent of each other because the former are mainly based on
the matter power spectrum suppression
while CGF arises from the squeezed limits of the 3pt correlation functions \cite{Ge:2023nnh}.
Being independent, the two PDFs
can simply multiply to obtain the combined PDF
which are shown as dash-dotted line in the left panel of \gfig{fig:pdf_DESI_DR1}.
The combined PDF decreases faster than either the
existing matter clustering or CGF ones.
In other words, combining these two methods can significantly enhance the
sensitivity of measuring the neutrino mass as well as the NMO.
The combined PDF gives a more stringent 95\% region of the neutrino mass
sum,
$\sum m_\nu < 0.092\,{\rm eV}$ and $\sum m_\nu < 0.121\,{\rm eV}$ for
NO (with prior $\sum m_\nu > 0.059\,{\rm eV}$) and IO (with prior $\sum m_\nu > 0.101\,{\rm eV}$), respectively.

With multiplication,
the combined relative probabilities for the NO and IO scenarios
scale as
$ P_{\rm NO} \propto P_{\rm NO, DESI} \times P_{\rm NO, CGF}$
and
$ P_{\rm IO} \propto P_{\rm IO, DESI} \times P_{\rm IO, CGF}$
with the relative probabilities in \geqn{eq:PNOPIO} and \geqn{eq:PNO_IO_CGF}.
After normalization,
the relative probabilities for NO and IO become,
\begin{align}
  P_{\rm NO} \approx 98.2\%,
\quad {\rm and} \quad
  P_{\rm IO} \approx 1.8\%.
\label{eq:PNOPIO_combined}
\end{align}
Thus, there is a 2.37\,$\sigma$ preference of NO 
which is greatly improved from
the current 1.64\,$\sigma$ preference of NO (as discussed around \geqn{eq:PNOPIO}).

The combined PDFs are
shown with dash-dotted lines in \gfig{fig:pdf_DESI_DR1}.
Using the relative probabilities for NO and IO in \geqn{eq:PNOPIO_combined},
we can also sample the distribution of the lightest neutrino mass as 
dash-dotted lines in \gfig{fig:pdf_DESI_DR1}.
The maximum probability also occurs around $0.01\,{\rm eV}$ for both NO and IO.

\section{The Effect of CGF Measurements on Beta Decay Experiments}
\label{sec:CGF_TE}

As shown in the previous section, 
the projected CGF observation at DESI can significantly improve
the cosmological sensitivity of the neutrino mass and ordering.
Since the beta decay observables are essentially the neutrino masses, one
would expect the CGF measurements to also play a pivotal role
in the terrestrial beta decay experiments.
Notably, both the neutrinoless
double beta decay and the single beta decay are considered.
In this section, we explore the synergy between the
cosmological measurements and the
two types of terrestrial beta decay experiments.

\subsection{Neutrinoless double beta decay}

If neutrinos are of the Majorana type,
the neutrinoless double beta decay ($0\nu\beta\beta$) can occur.
Its transition probability is
regulated by the effective Majorana mass,
\begin{align} 
  m_{\beta \beta}
\equiv
  \sum U_{e k}^2 m_k,
\end{align}
where $U_{e k}$ is a matrix element of the
PMNS matrix.
Connecting the mass and flavor eigenstates, $\nu_\alpha = U_{\alpha i} \nu_i$, the PMNS matrix can be
parameterized as \cite{Ge:2016tfx,Ge:2019ldu},
\begin{align}
  U
\equiv
\left(\begin{array}{ccc}
c_s c_r & s_s c_r & s_r e^{-i \delta_D} \\
-c_a s_s-s_a s_r c_s e^{i \delta_D} & c_a c_s-s_a s_r s_s e^{i \delta_D} & s_a c_r \\
s_a s_s-c_a s_r c_s e^{i \delta_D} & -s_a c_s-c_a s_r s_s e^{i \delta_D} & c_a c_r
\end{array}\right) 
  \mathcal{Q},
\label{eq:def_PMNS}
\end{align}
where the mixing angles $\theta_{ij}$ and mass squared differences
$\Delta m_{ij}^2$ are denoted,
\begin{align}
  \theta_a \equiv \theta_{23}, 
\quad 
  \theta_r \equiv \theta_{13}, 
\quad 
  \theta_s \equiv \theta_{12}, 
\quad 
  \Delta m_a^2 \equiv \Delta m_{13}^2, 
\quad 
  \Delta m_s^2 \equiv \Delta m_{12}^2,
\end{align} 
according to their roles in the atmospheric ($a$),
reactor ($r$), and solar ($s$) neutrino oscillations.
The diagonal rephasing matrix 
$\mathcal Q \equiv {\rm diag} \{ e^{-i \delta_{\rm M1}/2 }, 1, e^{-i \delta_{\rm M3}/2 } \}$
contains two independent Majorana CP phases.
In the context of NO, the expression for the effective Majorana mass $m_{\beta \beta}$ is presented as,
\begin{align}
  m_{\beta \beta}
=
  m_1 c_r^2 c_s^2 e^{i \delta_{\rm M  1}}
+
  \sqrt{m_1^2+ m_s^2} c_r^2 s_s^2
+
  \sqrt{m_1^2+ m_a^2} s_r^2 e^{i \delta_{\rm M  3}},
\label{eq:mee_NO}
\end{align}
Accordingly, in the case of IO, the expression for the effective mass 
$m_{\beta \beta}$ is given by,
\begin{align}
  m_{\beta \beta}
=
  \sqrt{ m_3^2 + m_a^2 } c_r^2 c_s^2 e^{i \delta_{\rm M  1}}
+
  \sqrt{ m_3^2 + m_a^2 + m_s^2} c_r^2 s_s^2
+
  m_3 s_r^2 e^{i \delta_{\rm M  3}}.
\label{eq:mee_IO}
\end{align}

\begin{figure}[t]
\centering
\includegraphics[width=0.49\textwidth]{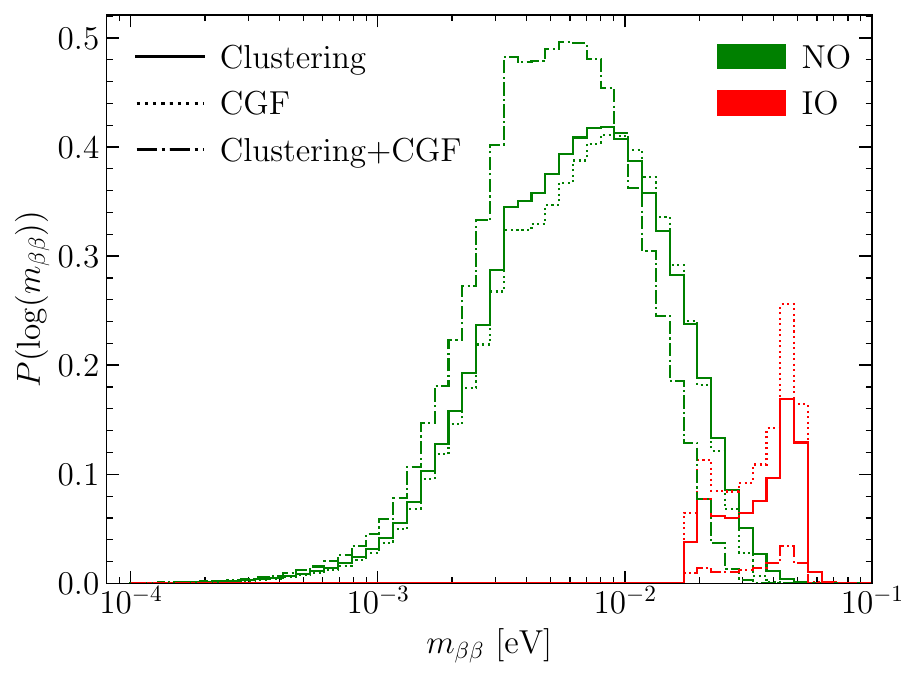}
\includegraphics[width=0.49\textwidth]{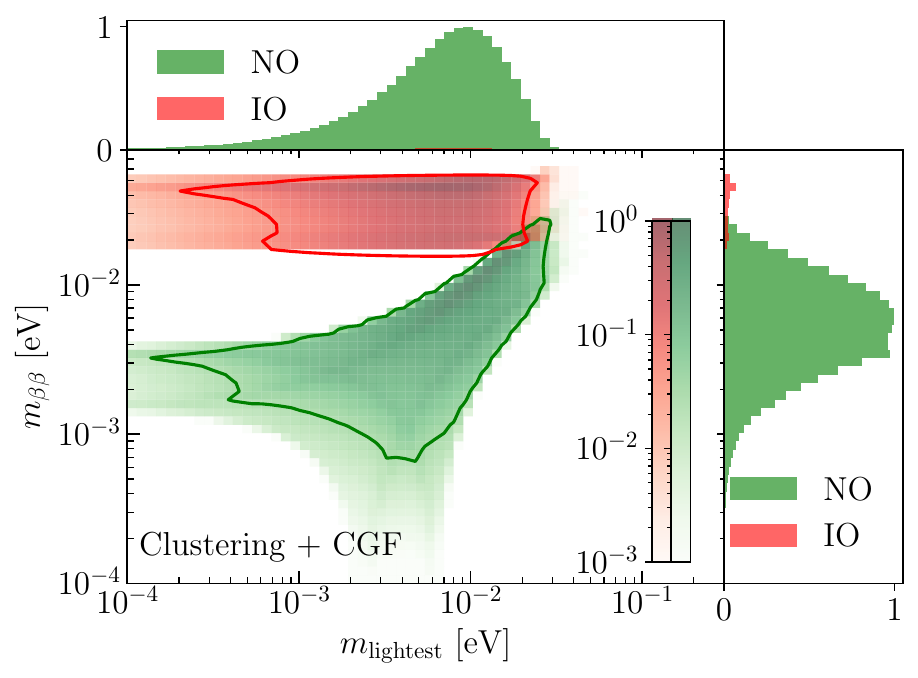}
\caption{
{\bf Left:}
The distribution of the effective Majorana mass $m_{\beta \beta}$
for NO (green) and IO (red) using the PDFs of
the existing matter clustering methods (solid), CGF (dotted),
and their combination (dash-dotted).
{\bf Right:} 
The 2D contour plot for the effective Majorana mass 
$m_{\beta \beta}$ as a function of the lightest neutrino mass $m_{1,3}$,
considering the combined constraint with both the existing matter clustering methods and CGF.
The contour with solid line encloses the data 
within the 95\%\,C.L. for both NO (green) and IO (red).
The projection of this contour along
the effective mass $m_{\beta \beta}$ on the right
(the lightest neutrino mass $m_{\rm lightest}$ at the top)
is the same as
the dash-dotted line in the left panel (the right panel of \protect\gfig{fig:pdf_DESI_DR1})}.
\label{fig:contour_mee_m1_combine}
\end{figure}

We then sample the effective Majorana mass $m_{\beta \beta}$
distribution as shown in the left panel of \gfig{fig:contour_mee_m1_combine}.
The mixing angles and mass squared differences considered in this analysis are
as follows: $\theta_r = 8.5^{\circ} \pm 0.2^{\circ}$,
$\theta_s = 33.48^{\circ} \pm 0.76^{\circ}$,
$|\Delta m_a^2| \approx 2.47 \times 10^{-3}\,{\rm eV}^2$,
and $\Delta m_s^2 \approx 7.54 \times 10^{-5}\,{\rm eV}^2$ \cite{ParticleDataGroup:2022pth}.
Without any prior assumption on the Majorana CP phases,
$\delta_{\rm M1}$ and $\delta_{\rm M3}$ 
are uniformly distributed in the whole range $[0, 2\pi]$.
The neutrino mass values are regulated by the PDFs from
the existing matter clustering methods, CGF,
and their combination as weighted by the corresponding
relative probabilities in \geqn{eq:PNOPIO}, \geqn{eq:PNO_IO_CGF}, and \geqn{eq:PNOPIO_combined}.
While the effective Majorana mass $m_{\beta \beta}$ is in
the $\mathcal O(10)$\,meV range for IO, its counterpart
for NO is mainly
in the $\mathcal O(1)$\,meV region. Between NO and IO, their
overlap happens in the IO range of $\mathcal O(10)$\,meV.
Combining the existing matter clustering and CGF
constraints decreases the probability of IO
and enhances its NO counterpart as discussed
in \gsec{sec:CGF_combine}.
At the same time, it also reduces the overlap between NO and IO
for the effective Majorana mass $m_{\beta \beta}$.

For NO (green), the maximum probabilities appear at $10\,{\rm meV}$
for both cases with either the existing matter clustering methods or CGF
which is a reflection of the fact that they have roughly the same
sensitivity on the neutrino mass measurement. However, the combination
of the existing matter clustering and CGF methods shifts
the maximum probability to
$m_{\beta \beta} = (3 \sim 7)\,{\rm meV}$. 
These values are below the limits $m_{\beta \beta} \lesssim (36 \sim 156)\,{\rm meV}$
and $m_{\beta \beta} \lesssim (79 \sim 180)\,{\rm meV}$
set by the current experiments,
KamLAND-Zen \cite{KamLAND-Zen:2022tow} and GERDA \cite{GERDA:2020xhi},
respectively. 
The future neutrinoless double-beta decay experiments
aim to further narrow this range to
$m_{\beta \beta} = (18.4 \pm 1.3)\,{\rm meV}$
\cite{Agostini:2021kba} and a target region of
$m_{\beta \beta} \approx (8 \sim 10)\,{\rm meV}$ \cite{Agostini:2022zub},
which is close to the regions of $(3 \sim 7)\,{\rm meV}$ preferred in our analysis.
To reach this,
it might take ten years with 100 tone ${}^{76}{\rm Ge}$ \cite{Mei:2024kvs}.
Once reaching the meV mass sensitivity, it might be possible to determine
the two Majorana CP phases simultaneously \cite{Ge:2016tfx,Cao:2019hli}
and testify the Dark-LMA (dark large mixing angle) solution \cite{Ge:2019ldu}.

For the IO (red) scenario, the $m_{\beta \beta}$ distribution exhibits
two distinct peaks around $20\,{\rm meV}$ and $50\,{\rm meV}$,
which are prominent for both the existing matter clustering and CGF cases as shown in the left panel of
\gfig{fig:contour_mee_m1_combine}.
This is because we assume a flat distribution in the two Majorana CP phases.
Since $m_{\beta \beta}$ is mainly a cosine function of the Majorana CP phases, its values concentrate at the extreme values of the cosine function
that lies at the boundaries of the IO band.
These peaks fall within the sensitivity range of the ongoing and
future neutrinoless double-beta decay experiments as mentioned above.

We plot the two-dimensional (2D) contour of 
$m_{\rm lightest}$--$m_{\beta \beta}$
in the right panel of \gfig{fig:contour_mee_m1_combine},
using the lightest neutrino mass and effective Majorana mass distributions
in \gfig{fig:pdf_DESI_DR1} and \gfig{fig:contour_mee_m1_combine}.
We only present the combined results of the
existing matter clustering methods and CGF
as the separate cases yield similar results.
For NO in the 2D contour, there is a funnel region of
$1\,\mbox{meV} < m_1 < 10\,\mbox{meV}$.
In this region, $m_{\beta \beta}$ is very small
which poses a significant challenge for the
neutrinoless double beta decay experiments. Fortunately, this funnel region
lies outside the maximum probability area, namely the 95\%\,C.L.
enclosed contour. For IO, a bimodal structure also appears in the 2D contour,
with the maximum probability located at the boundaries of the 95\%\,C.L.
region.
\footnote{
For the IO case, the $m_{\beta \beta}$ distribution drops off rapidly around $m_{\beta \beta} \approx 0.02\,{\rm eV}$, based on the binning we used. As a result, the 95\%\,C.L. contour lies slightly below the colored region.
}

\subsection{Single beta decay}

The single beta decay also serves as a crucial
terrestrial experiment for measuring the neutrino masses.
Especially, the endpoint of the beta spectrum directly correlates with the neutrino mass.
For the three-flavor neutrino mixing, the deviation
from the predicted electron spectrum with massless neutrino
around the endpoint can be effectively parameterized in terms of
a combination of the neutrino masses as,
\begin{align} 
  m_\beta^2 
\equiv
  \sum_k |U_{e k}|^2  m_k^2,
\end{align}
where $U_{e k}$ is the first row of the PMNS matrix in \geqn{eq:def_PMNS}.
Depending on the NMO,
the effective mass $m_\beta$ can be expressed as,
\begin{align}
  m_\beta
=
\begin{cases}
    \sqrt{
  m_1^2 c_r^2 c_s^2
+
  (m_1^2+ \Delta m_s^2) c_r^2 s_s^2
+
  (m_1^2+ |\Delta m_a^2|) s_r^2}
& \quad \mbox{NO},
\\[1mm]
\sqrt{
  (m_3^2 + |\Delta m_a^2|) c_r^2 c_s^2
+
  (m_3^2 + |\Delta m_a^2| + \Delta m_s^2) c_r^2 s_s^2
+
  m_3^2 s_r^2}
& \quad \mbox{IO}.
\end{cases}
\label{eq:mbeta}
\end{align}

\begin{figure}[t]
\centering
\includegraphics[width=0.49\textwidth]{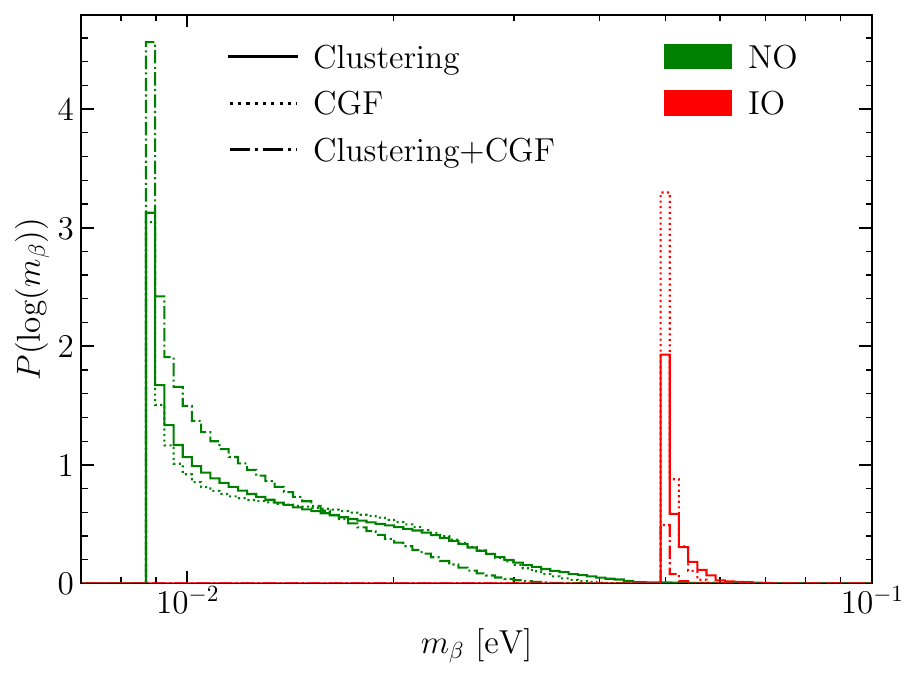}
\includegraphics[width=0.49\textwidth]{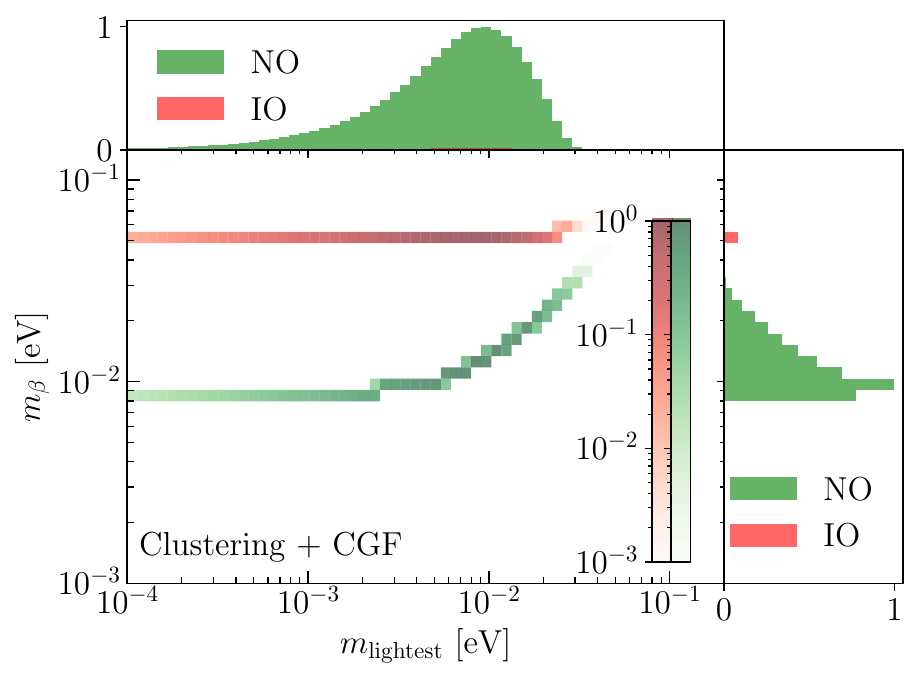}
\caption{
{\bf Left:}
The beta decay effective mass $m_{\beta}$ distribution for
the existing matter clustering methods, CGF, and
their combination for NO (green) and IO (red), respectively.
{\bf Right:} 
The contour for the
effective mass $m_{\beta}$ and
the lightest neutrino mass, considering the combined
measurements with both the existing matter clustering methods
and CGF. Its projections along the lightest neutrino mass
and the beta decay effective mass $m_{\beta}$ are
shown in the top and right sub-figures, respectively.
}
\label{fig:contour_mbeta_m1_combine}
\end{figure}

The distributions of the effective mass $m_{\beta}$
in the left panel of \gfig{fig:contour_mbeta_m1_combine}
are sampled using the PDFs of the
existing matter clustering methods, CGF,
and their combination, weighted by the corresponding
relative probabilities 
in \geqn{eq:PNOPIO}, \geqn{eq:PNO_IO_CGF}, and \geqn{eq:PNOPIO_combined}. 
Although the distributions for these three cases are quite similar,
the areas below the curves differ which
reflects the relative probabilities
between NO and IO as discussed in \gsec{sec:nu_ordering}.
Since the cosmological data prefer the lightest neutrino mass around
$m_{\rm lightest} \approx 0.01\,{\rm eV}$, the effective mass $m_\beta$
in \geqn{eq:mbeta} is approximately 
$8.7\,{\rm meV}$ (49\,{\rm meV}) for NO (IO), respectively.
These values correspond to the maximum probability in the left panel of 
\gfig{fig:contour_mbeta_m1_combine} for NO (green) and IO (red).
For the combined case of the existing matter clustering methods and CGF, the NO has a
larger probability concentrated around 
its maximum value of $m_{\beta} \approx 8.7\,{\rm meV}$
in comparison with the individual case of matter
clustering or CGF.

The 2D $m_\beta$--$m_{\rm lightest}$
contours are also presented 
in the right panel of \gfig{fig:contour_mbeta_m1_combine},
using the distributions of $m_{\rm lightest}$ and
$m_\beta$ shown in 
\gfig{fig:pdf_DESI_DR1} and \gfig{fig:contour_mbeta_m1_combine}.
Here, we only present the combined case
since the three cases
(mass sum, CGF, and their combination) have
similar behaviors. For NO, the $m_\beta$ region spans
from $0.01\,{\rm eV}$ to $0.04\,{\rm eV}$.
This distribution demonstrates a nearly monotonic relationship in this range,
where one $m_{\rm lightest}$ value corresponds uniquely to one $m_\beta$ value.
Once the neutrino mass is measured from $\beta$ decay within this range,
the lightest neutrino mass can be simultaneously determined, and vice versa. 

For IO, $m_\beta$ is almost always around $0.05\,{\rm eV}$.
This value is far from the current KATRIN bound of 
$m_{\beta} < 0.45\,{\rm eV}$ \cite{KATRIN:2021uub,Katrin:2024tvg} and the near-future KATRIN target
of $m_{\beta} < 0.2\,{\rm eV}$ with more data from tritium beta decay \cite{KATRIN:2005fny}.
In addition, the future Project 8 is expected to probe this region,
reaching nearly $0.04\,{\rm eV}$ \cite{Project8:2014ivu}.

\section{Conclusions}
\label{sec:conclusion}

In this paper, we demonstrate that the CGF provides a complementary cosmological measurement 
of the neutrino masses $m_i$, due to its mass fourth-power dependence $m_i^4$ \cite{Ge:2023nnh,Okoli:2016vmd}.
The CGF alone can give a constraint of the mass sum,
$\sum m_i < 0.109$\,eV ($\sum m_i < 0.129$\,eV) for NO (IO) at 95\% C.L., 
which is comparable to the existing matter clustering constraint
from the DESI analysis \cite{DESI:2024mwx}
$\sum m_i < 0.113$\,eV ($\sum m_i < 0.145$\,eV) for NO (IO).
The combination of CGF and clustering yields a more stringent constraint on the neutrino mass sum,
$\sum m_i < 0.092$\,eV ($\sum m_i < 0.121$\,eV) for NO (IO).
Using the fact that total mass falling below IO threshold
allows the discrimination of IO,
the CGF can also help to identify the neutrino mass ordering,
as demonstrated for the first time, upon incorporating
the projected CGF sensitivity at DESI, 
the preference for NO would significantly improve
from $1.64\,\sigma$ to $2.37\,\sigma$
with a prior of $\sum m_i > 0.059\,{\rm eV}$.

Including CGF,
the cosmological measurement yields distinct targets for neutrinoless double and single beta decays.
The combination of clustering and CGF provides 
the peak location of the effective mass of $m_{\beta \beta}$ near $3\,{\rm meV}$ ($50\,{\rm meV}$) for NO (IO), 
which can be reached in future neutrinoless double decay experiments \cite{Agostini:2021kba,Agostini:2022zub}.
For the single beta decay, the effective mass $m_{\beta}$ 
prefers a peak location at $8.7\,{\rm meV}$ (49\,{\rm meV}) for NO (IO), respectively.
These values are far from the current beta decay experiments, 
but sensitive at the future beta decay experiments such as Project 8 \cite{Project8:2014ivu}.

\section{Note Added}

"During the preparation of this paper, we note that DESI has released Data Release 2 (DR2) \cite{DESI:2025ejh}, where the IO is further disflavored compared with the DR1 used in this paper."

\section{AKNOWLEDGEMENT}

The authors are supported by the National Natural Science
Foundation of China (12425506, 12375101, 12090060 and 12090064)
and the SJTU Double First Class start-up fund (WF220442604).
SFG is also an affiliate member of Kavli IPMU, University of Tokyo.

\bibliographystyle{utphysGe}
\bibliography{nuOrderingCGF}

\providecommand{\href}[2]{#2}\begingroup\raggedright\begin{thebibliography}{10}

\bibitem{Super-Kamiokande:1998kpq}
{\bfseries Super-Kamiokande} Collaboration, Y.~Fukuda {et~al.}, ``{\it
  {Evidence for oscillation of atmospheric neutrinos}},''
  \href{http://dx.doi.org/10.1103/PhysRevLett.81.1562}{{Phys. Rev. Lett.}
  {\bfseries 81} (1998) 1562--1567},
  [\href{http://arxiv.org/abs/hep-ex/9807003}{{\ttfamily
  arXiv:hep-ex/9807003}}].

\bibitem{SNO:2002tuh}
{\bfseries SNO} Collaboration, Q.~R. Ahmad {et~al.}, ``{\it {Direct evidence
  for neutrino flavor transformation from neutral current interactions in the
  Sudbury Neutrino Observatory}},''
  \href{http://dx.doi.org/10.1103/PhysRevLett.89.011301}{{Phys. Rev. Lett.}
  {\bfseries 89} (2002) 011301},
  [\href{http://arxiv.org/abs/nucl-ex/0204008}{{\ttfamily
  arXiv:nucl-ex/0204008}}].

\bibitem{Mohapatra:2005wg}
R.~N. Mohapatra {et~al.}, ``{\it {Theory of neutrinos: A White paper}},''
  \href{http://dx.doi.org/10.1088/0034-4885/70/11/R02}{{Rept. Prog. Phys.}
  {\bfseries 70} (2007) 1757--1867},
  [\href{http://arxiv.org/abs/hep-ph/0510213}{{\ttfamily
  arXiv:hep-ph/0510213}}].

\bibitem{Gonzalez-Garcia:2007dlo}
M.~C. Gonzalez-Garcia and Michele Maltoni, ``{\it {Phenomenology with Massive
  Neutrinos}},'' \href{http://dx.doi.org/10.1016/j.physrep.2007.12.004}{{Phys.
  Rept.} {\bfseries 460} (2008) 1--129},
  [\href{http://arxiv.org/abs/0704.1800}{{\ttfamily arXiv:0704.1800}}
  [hep-ph]].

\bibitem{Giganti:2017fhf}
Claudio Giganti, St\'ephane Lavignac, and Marco Zito, ``{\it {Neutrino
  oscillations: The rise of the PMNS paradigm}},''
  \href{http://dx.doi.org/10.1016/j.ppnp.2017.10.001}{{Prog. Part. Nucl. Phys.}
  {\bfseries 98} (2018) 1--54},
  [\href{http://arxiv.org/abs/1710.00715}{{\ttfamily arXiv:1710.00715}}
  [hep-ex]].

\bibitem{Esteban:2020cvm}
Ivan Esteban, M.~C. Gonzalez-Garcia, Michele Maltoni, Thomas Schwetz, and
  Albert Zhou, ``{\it {The fate of hints: updated global analysis of
  three-flavor neutrino oscillations}},''
  \href{http://dx.doi.org/10.1007/JHEP09(2020)178}{{JHEP} {\bfseries 09} (2020)
  178}, [\href{http://arxiv.org/abs/2007.14792}{{\ttfamily arXiv:2007.14792}}
  [hep-ph]].

\bibitem{deSalas:2020pgw}
P.~F. de~Salas, D.~V. Forero, S.~Gariazzo, P.~Mart\'\i{}nez-Mirav\'e, O.~Mena,
  C.~A. Ternes, M.~T\'ortola, and J.~W.~F. Valle, ``{\it {2020 global
  reassessment of the neutrino oscillation picture}},''
  \href{http://dx.doi.org/10.1007/JHEP02(2021)071}{{JHEP} {\bfseries 02} (2021)
  071}, [\href{http://arxiv.org/abs/2006.11237}{{\ttfamily arXiv:2006.11237}}
  [hep-ph]].

\bibitem{Capozzi:2021fjo}
Francesco Capozzi, Eleonora Di~Valentino, Eligio Lisi, Antonio Marrone,
  Alessandro Melchiorri, and Antonio Palazzo, ``{\it {Unfinished fabric of the
  three neutrino paradigm}},''
  \href{http://dx.doi.org/10.1103/PhysRevD.104.083031}{{Phys. Rev. D}
  {\bfseries 104} no.~8, (2021) 083031},
  [\href{http://arxiv.org/abs/2107.00532}{{\ttfamily arXiv:2107.00532}}
  [hep-ph]].

\bibitem{JUNO:2015zny}
{\bfseries JUNO} Collaboration, Fengpeng An {et~al.}, ``{\it {Neutrino Physics
  with JUNO}},'' \href{http://dx.doi.org/10.1088/0954-3899/43/3/030401}{{J.
  Phys. G} {\bfseries 43} no.~3, (2016) 030401},
  [\href{http://arxiv.org/abs/1507.05613}{{\ttfamily arXiv:1507.05613}}
  [physics.ins-det]].

\bibitem{JUNO:2015sjr}
{\bfseries JUNO} Collaboration, Zelimir Djurcic {et~al.}, ``{\it {JUNO
  Conceptual Design Report}},''
  [\href{http://arxiv.org/abs/1508.07166}{{\ttfamily arXiv:1508.07166}}
  [physics.ins-det]].

\bibitem{Zhang:2021adu}
{\bfseries JUNO} Collaboration, Jinnan Zhang, ``{\it {JUNO Oscillation
  Physics}},'' \href{http://dx.doi.org/10.1088/1742-6596/2156/1/012110}{{J.
  Phys. Conf. Ser.} {\bfseries 2156} no.~1, (2021) 012110},
  [\href{http://arxiv.org/abs/2111.10112}{{\ttfamily arXiv:2111.10112}}
  [physics.ins-det]].

\bibitem{ParticleDataGroup:2022pth}
{\bfseries Particle Data Group} Collaboration, R.~L. Workman {et~al.}, ``{\it
  {Review of Particle Physics}},''
  \href{http://dx.doi.org/10.1093/ptep/ptac097}{{PTEP} {\bfseries 2022} (2022)
  083C01}.

\bibitem{KATRIN:2021uub}
{\bfseries KATRIN} Collaboration, M.~Aker {et~al.}, ``{\it {Direct
  neutrino-mass measurement with sub-electronvolt sensitivity}},''
  \href{http://dx.doi.org/10.1038/s41567-021-01463-1}{{Nature Phys.} {\bfseries
  18} no.~2, (2022) 160--166},
  [\href{http://arxiv.org/abs/2105.08533}{{\ttfamily arXiv:2105.08533}}
  [hep-ex]].

\bibitem{Katrin:2024tvg}
{\bfseries Katrin} Collaboration, M.~Aker {et~al.}, ``{\it {Direct
  neutrino-mass measurement based on 259 days of KATRIN data}},''
  [\href{http://arxiv.org/abs/2406.13516}{{\ttfamily arXiv:2406.13516}}
  [nucl-ex]].

\bibitem{KATRIN:2005fny}
{\bfseries KATRIN} Collaboration, J.~Angrik {et~al.}, ``{\it {KATRIN design
  report 2004}},''.

\bibitem{Project8:2014ivu}
{\bfseries Project 8} Collaboration, D.~M. Asner {et~al.}, ``{\it {Single
  electron detection and spectroscopy via relativistic cyclotron radiation}},''
  \href{http://dx.doi.org/10.1103/PhysRevLett.114.162501}{{Phys. Rev. Lett.}
  {\bfseries 114} no.~16, (2015) 162501},
  [\href{http://arxiv.org/abs/1408.5362}{{\ttfamily arXiv:1408.5362}}
  [physics.ins-det]].

\bibitem{KamLAND-Zen:2022tow}
{\bfseries KamLAND-Zen} Collaboration, S.~Abe {et~al.}, ``{\it {Search for the
  Majorana Nature of Neutrinos in the Inverted Mass Ordering Region with
  KamLAND-Zen}},''
  \href{http://dx.doi.org/10.1103/PhysRevLett.130.051801}{{Phys. Rev. Lett.}
  {\bfseries 130} no.~5, (2023) 051801},
  [\href{http://arxiv.org/abs/2203.02139}{{\ttfamily arXiv:2203.02139}}
  [hep-ex]].

\bibitem{GERDA:2020xhi}
{\bfseries GERDA} Collaboration, M.~Agostini {et~al.}, ``{\it {Final Results of
  GERDA on the Search for Neutrinoless Double-$\beta$ Decay}},''
  \href{http://dx.doi.org/10.1103/PhysRevLett.125.252502}{{Phys. Rev. Lett.}
  {\bfseries 125} no.~25, (2020) 252502},
  [\href{http://arxiv.org/abs/2009.06079}{{\ttfamily arXiv:2009.06079}}
  [nucl-ex]].

\bibitem{Agostini:2021kba}
Matteo Agostini, Giovanni Benato, Jason~A. Detwiler, Javier Men\'endez, and
  Francesco Vissani, ``{\it {Testing the inverted neutrino mass ordering with
  neutrinoless double-\ensuremath{\beta} decay}},''
  \href{http://dx.doi.org/10.1103/PhysRevC.104.L042501}{{Phys. Rev. C}
  {\bfseries 104} no.~4, (2021) L042501},
  [\href{http://arxiv.org/abs/2107.09104}{{\ttfamily arXiv:2107.09104}}
  [hep-ph]].

\bibitem{Lesgourgues:2006nd}
Julien Lesgourgues and Sergio Pastor, ``{\it {Massive neutrinos and
  cosmology}},'' \href{http://dx.doi.org/10.1016/j.physrep.2006.04.001}{{Phys.
  Rept.} {\bfseries 429} (2006) 307--379},
  [\href{http://arxiv.org/abs/astro-ph/0603494}{{\ttfamily
  arXiv:astro-ph/0603494}}].

\bibitem{TopicalConvenersKNAbazajianJECarlstromATLee:2013bxd}
{\bfseries Topical Conveners: K.N. Abazajian, J.E. Carlstrom, A.T. Lee}
  Collaboration, K.~N. Abazajian {et~al.}, ``{\it {Neutrino Physics from the
  Cosmic Microwave Background and Large Scale Structure}},''
  \href{http://dx.doi.org/10.1016/j.astropartphys.2014.05.014}{{Astropart.
  Phys.} {\bfseries 63} (2015) 66--80},
  [\href{http://arxiv.org/abs/1309.5383}{{\ttfamily arXiv:1309.5383}}
  [astro-ph.CO]].

\bibitem{Dvorkin:2019jgs}
Cora Dvorkin {et~al.}, ``{\it {Neutrino Mass from Cosmology: Probing Physics
  Beyond the Standard Model}},''
  [\href{http://arxiv.org/abs/1903.03689}{{\ttfamily arXiv:1903.03689}}
  [astro-ph.CO]].

\bibitem{DiValentino:2024xsv}
Eleonora Di~Valentino, Stefano Gariazzo, and Olga Mena, ``{\it {Neutrinos in
  Cosmology}},'' [\href{http://arxiv.org/abs/2404.19322}{{\ttfamily
  arXiv:2404.19322}} [astro-ph.CO]].

\bibitem{Planck:2018vyg}
{\bfseries Planck} Collaboration, N.~Aghanim {et~al.}, ``{\it {Planck 2018
  results. VI. Cosmological parameters}},''
  \href{http://dx.doi.org/10.1051/0004-6361/201833910}{{Astron. Astrophys.}
  {\bfseries 641} (2020) A6},
  [\href{http://arxiv.org/abs/1807.06209}{{\ttfamily arXiv:1807.06209}}
  [astro-ph.CO]]. [Erratum: Astron.Astrophys. 652, C4 (2021)].

\bibitem{DESI:2024mwx}
{\bfseries DESI} Collaboration, A.~G. Adame {et~al.}, ``{\it {DESI 2024 VI:
  Cosmological Constraints from the Measurements of Baryon Acoustic
  Oscillations}},'' [\href{http://arxiv.org/abs/2404.03002}{{\ttfamily
  arXiv:2404.03002}} [astro-ph.CO]].

\bibitem{Hannestad:2016fog}
Steen Hannestad and Thomas Schwetz, ``{\it {Cosmology and the neutrino mass
  ordering}},'' \href{http://dx.doi.org/10.1088/1475-7516/2016/11/035}{{JCAP}
  {\bfseries 11} (2016) 035},
  [\href{http://arxiv.org/abs/1606.04691}{{\ttfamily arXiv:1606.04691}}
  [astro-ph.CO]].

\bibitem{Vagnozzi:2017ovm}
Sunny Vagnozzi, Elena Giusarma, Olga Mena, Katherine Freese, Martina Gerbino,
  Shirley Ho, and Massimiliano Lattanzi, ``{\it {Unveiling $\nu$ secrets with
  cosmological data: neutrino masses and mass hierarchy}},''
  \href{http://dx.doi.org/10.1103/PhysRevD.96.123503}{{Phys. Rev. D} {\bfseries
  96} no.~12, (2017) 123503},
  [\href{http://arxiv.org/abs/1701.08172}{{\ttfamily arXiv:1701.08172}}
  [astro-ph.CO]].

\bibitem{Tanseri:2022zfe}
Isabelle Tanseri, Steffen Hagstotz, Sunny Vagnozzi, Elena Giusarma, and
  Katherine Freese, ``{\it {Updated neutrino mass constraints from galaxy
  clustering and CMB lensing-galaxy cross-correlation measurements}},''
  \href{http://dx.doi.org/10.1016/j.jheap.2022.07.002}{{JHEAp} {\bfseries 36}
  (2022) 1--26}, [\href{http://arxiv.org/abs/2207.01913}{{\ttfamily
  arXiv:2207.01913}} [astro-ph.CO]].

\bibitem{Craig:2024tky}
Nathaniel Craig, Daniel Green, Joel Meyers, and Surjeet Rajendran, ``{\it {No
  $\nu$s is Good News}},'' [\href{http://arxiv.org/abs/2405.00836}{{\ttfamily
  arXiv:2405.00836}} [astro-ph.CO]].

\bibitem{Green:2024xbb}
Daniel Green and Joel Meyers, ``{\it {The Cosmological Preference for Negative
  Neutrino Mass}},'' [\href{http://arxiv.org/abs/2407.07878}{{\ttfamily
  arXiv:2407.07878}} [astro-ph.CO]].

\bibitem{Herold:2024enb}
Laura Herold, Elisa G.~M. Ferreira, and Lukas Heinrich, ``{\it {Profile
  Likelihoods in Cosmology: When, Why and How illustrated with $\Lambda$CDM,
  Massive Neutrinos and Dark Energy}},''
  [\href{http://arxiv.org/abs/2408.07700}{{\ttfamily arXiv:2408.07700}}
  [astro-ph.CO]].

\bibitem{Wang:2024hen}
Deng Wang, Olga Mena, Eleonora Di~Valentino, and Stefano Gariazzo, ``{\it
  {Updating neutrino mass constraints with Background measurements}},''
  [\href{http://arxiv.org/abs/2405.03368}{{\ttfamily arXiv:2405.03368}}
  [astro-ph.CO]].

\bibitem{Jiang:2024viw}
Jun-Qian Jiang, William Giar\`e, Stefano Gariazzo, Maria~Giovanna Dainotti,
  Eleonora Di~Valentino, Olga Mena, Davide Pedrotti, Simony~Santos da~Costa,
  and Sunny Vagnozzi, ``{\it {Neutrino cosmology after DESI: tightest mass
  upper limits, preference for the normal ordering, and tension with
  terrestrial observations}},''
  [\href{http://arxiv.org/abs/2407.18047}{{\ttfamily arXiv:2407.18047}}
  [astro-ph.CO]].

\bibitem{RoyChoudhury:2019hls}
Shouvik Roy~Choudhury and Steen Hannestad, ``{\it {Updated results on neutrino
  mass and mass hierarchy from cosmology with Planck 2018 likelihoods}},''
  \href{http://dx.doi.org/10.1088/1475-7516/2020/07/037}{{JCAP} {\bfseries 07}
  (2020) 037}, [\href{http://arxiv.org/abs/1907.12598}{{\ttfamily
  arXiv:1907.12598}} [astro-ph.CO]].

\bibitem{Motloch:2019gux}
Pavel Motloch and Wayne Hu, ``{\it {Lensinglike tensions in the $Planck$ legacy
  release}},'' \href{http://dx.doi.org/10.1103/PhysRevD.101.083515}{{Phys. Rev.
  D} {\bfseries 101} no.~8, (2020) 083515},
  [\href{http://arxiv.org/abs/1912.06601}{{\ttfamily arXiv:1912.06601}}
  [astro-ph.CO]].

\bibitem{DiValentino:2021imh}
Eleonora Di~Valentino and Alessandro Melchiorri, ``{\it {Neutrino Mass Bounds
  in the Era of Tension Cosmology}},''
  \href{http://dx.doi.org/10.3847/2041-8213/ac6ef5}{{Astrophys. J. Lett.}
  {\bfseries 931} no.~2, (2022) L18},
  [\href{http://arxiv.org/abs/2112.02993}{{\ttfamily arXiv:2112.02993}}
  [astro-ph.CO]].

\bibitem{Allali:2024aiv}
Itamar~J. Allali and Alessio Notari, ``{\it {Neutrino mass bounds from DESI
  2024 are relaxed by Planck PR4 and cosmological supernovae}},''
  [\href{http://arxiv.org/abs/2406.14554}{{\ttfamily arXiv:2406.14554}}
  [astro-ph.CO]].

\bibitem{Naredo-Tuero:2024sgf}
Daniel Naredo-Tuero, Miguel Escudero, Enrique Fern\'andez-Mart\'\i{}nez, Xabier
  Marcano, and Vivian Poulin, ``{\it {Living at the Edge: A Critical Look at
  the Cosmological Neutrino Mass Bound}},''
  [\href{http://arxiv.org/abs/2407.13831}{{\ttfamily arXiv:2407.13831}}
  [astro-ph.CO]].

\bibitem{Yadav:2024duq}
Anita Yadav, Suresh Kumar, Cihad Kibris, and Ozgur Akarsu, ``{\it
  {$\Lambda_{\rm s}$CDM cosmology: Alleviating major cosmological tensions by
  predicting standard neutrino properties}},''
  [\href{http://arxiv.org/abs/2406.18496}{{\ttfamily arXiv:2406.18496}}
  [astro-ph.CO]].

\bibitem{Du:2024pai}
Guo-Hong Du, Peng-Ju Wu, Tian-Nuo Li, and Xin Zhang, ``{\it {Impacts of dark
  energy on weighing neutrinos after DESI BAO}},''
  [\href{http://arxiv.org/abs/2407.15640}{{\ttfamily arXiv:2407.15640}}
  [astro-ph.CO]].

\bibitem{Elbers:2024sha}
Willem Elbers, Carlos~S. Frenk, Adrian Jenkins, Baojiu Li, and Silvia Pascoli,
  ``{\it {Negative neutrino masses as a mirage of dark energy}},''
  [\href{http://arxiv.org/abs/2407.10965}{{\ttfamily arXiv:2407.10965}}
  [astro-ph.CO]].

\bibitem{Reboucas:2024smm}
Jo\~ao Rebou\c{c}as, Diogo H.~F. de~Souza, Kunhao Zhong, Vivian Miranda, and
  Rogerio Rosenfeld, ``{\it {Investigating Late-Time Dark Energy and Massive
  Neutrinos in Light of DESI Y1 BAO}},''
  [\href{http://arxiv.org/abs/2408.14628}{{\ttfamily arXiv:2408.14628}}
  [astro-ph.CO]].

\bibitem{Shao:2024mag}
Helen Shao, Jahmour~J. Givans, Jo~Dunkley, Mathew Madhavacheril, Frank Qu,
  Gerrit Farren, and Blake Sherwin, ``{\it {Cosmological limits on the neutrino
  mass sum for beyond-$\Lambda$CDM models}},''
  [\href{http://arxiv.org/abs/2409.02295}{{\ttfamily arXiv:2409.02295}}
  [astro-ph.CO]].

\bibitem{Sen:2024pgb}
Manibrata Sen and Alexei~Y. Smirnov, ``{\it {Neutrinos with refractive masses
  and the DESI BAO results}},''
  [\href{http://arxiv.org/abs/2407.02462}{{\ttfamily arXiv:2407.02462}}
  [hep-ph]].

\bibitem{Forconi:2023akg}
Matteo Forconi, Eleonora Di~Valentino, Alessandro Melchiorri, and Supriya Pan,
  ``{\it {Possible impact of non-Gaussianities on cosmological constraints in
  neutrino physics}},''
  \href{http://dx.doi.org/10.1103/PhysRevD.109.123532}{{Phys. Rev. D}
  {\bfseries 109} no.~12, (2024) 123532},
  [\href{http://arxiv.org/abs/2311.04038}{{\ttfamily arXiv:2311.04038}}
  [astro-ph.CO]].

\bibitem{Bellomo:2016xhl}
Nicola Bellomo, Emilio Bellini, Bin Hu, Raul Jimenez, Carlos Pena-Garay, and
  Licia Verde, ``{\it {Hiding neutrino mass in modified gravity
  cosmologies}},''
  \href{http://dx.doi.org/10.1088/1475-7516/2017/02/043}{{JCAP} {\bfseries 02}
  (2017) 043}, [\href{http://arxiv.org/abs/1612.02598}{{\ttfamily
  arXiv:1612.02598}} [astro-ph.CO]].

\bibitem{Lorenz:2018fzb}
Christiane~S. Lorenz, Lena Funcke, Erminia Calabrese, and Steen Hannestad,
  ``{\it {Time-varying neutrino mass from a supercooled phase transition:
  current cosmological constraints and impact on the $\Omega_m$-$\sigma_8$
  plane}},'' \href{http://dx.doi.org/10.1103/PhysRevD.99.023501}{{Phys. Rev. D}
  {\bfseries 99} no.~2, (2019) 023501},
  [\href{http://arxiv.org/abs/1811.01991}{{\ttfamily arXiv:1811.01991}}
  [astro-ph.CO]].

\bibitem{Oldengott:2019lke}
Isabel~M. Oldengott, Gabriela Barenboim, Sarah Kahlen, Jordi Salvado, and
  Dominik~J. Schwarz, ``{\it {How to relax the cosmological neutrino mass
  bound}},'' \href{http://dx.doi.org/10.1088/1475-7516/2019/04/049}{{JCAP}
  {\bfseries 04} (2019) 049},
  [\href{http://arxiv.org/abs/1901.04352}{{\ttfamily arXiv:1901.04352}}
  [astro-ph.CO]].

\bibitem{Alvey:2021sji}
James Alvey, Miguel Escudero, and Nashwan Sabti, ``{\it {What can CMB
  observations tell us about the neutrino distribution function?}},''
  \href{http://dx.doi.org/10.1088/1475-7516/2022/02/037}{{JCAP} {\bfseries 02}
  no.~02, (2022) 037}, [\href{http://arxiv.org/abs/2111.12726}{{\ttfamily
  arXiv:2111.12726}} [astro-ph.CO]].

\bibitem{Beacom:2004yd}
John~F. Beacom, Nicole~F. Bell, and Scott Dodelson, ``{\it {Neutrinoless
  universe}},'' \href{http://dx.doi.org/10.1103/PhysRevLett.93.121302}{{Phys.
  Rev. Lett.} {\bfseries 93} (2004) 121302},
  [\href{http://arxiv.org/abs/astro-ph/0404585}{{\ttfamily
  arXiv:astro-ph/0404585}}].

\bibitem{Farzan:2015pca}
Yasaman Farzan and Steen Hannestad, ``{\it {Neutrinos secretly converting to
  lighter particles to please both KATRIN and the cosmos}},''
  \href{http://dx.doi.org/10.1088/1475-7516/2016/02/058}{{JCAP} {\bfseries 02}
  (2016) 058}, [\href{http://arxiv.org/abs/1510.02201}{{\ttfamily
  arXiv:1510.02201}} [hep-ph]].

\bibitem{Chacko:2019nej}
Zackaria Chacko, Abhish Dev, Peizhi Du, Vivian Poulin, and Yuhsin Tsai, ``{\it
  {Cosmological Limits on the Neutrino Mass and Lifetime}},''
  \href{http://dx.doi.org/10.1007/JHEP04(2020)020}{{JHEP} {\bfseries 04} (2020)
  020}, [\href{http://arxiv.org/abs/1909.05275}{{\ttfamily arXiv:1909.05275}}
  [hep-ph]].

\bibitem{Chacko:2020hmh}
Zackaria Chacko, Abhish Dev, Peizhi Du, Vivian Poulin, and Yuhsin Tsai, ``{\it
  {Determining the Neutrino Lifetime from Cosmology}},''
  \href{http://dx.doi.org/10.1103/PhysRevD.103.043519}{{Phys. Rev. D}
  {\bfseries 103} no.~4, (2021) 043519},
  [\href{http://arxiv.org/abs/2002.08401}{{\ttfamily arXiv:2002.08401}}
  [astro-ph.CO]].

\bibitem{Escudero:2020ped}
Miguel Escudero, Jacobo Lopez-Pavon, Nuria Rius, and Stefan Sandner, ``{\it
  {Relaxing Cosmological Neutrino Mass Bounds with Unstable Neutrinos}},''
  \href{http://dx.doi.org/10.1007/JHEP12(2020)119}{{JHEP} {\bfseries 12} (2020)
  119}, [\href{http://arxiv.org/abs/2007.04994}{{\ttfamily arXiv:2007.04994}}
  [hep-ph]].

\bibitem{Escudero:2022gez}
Miguel Escudero, Thomas Schwetz, and Jorge Terol-Calvo, ``{\it {A seesaw model
  for large neutrino masses in concordance with cosmology}},''
  \href{http://dx.doi.org/10.1007/JHEP02(2023)142}{{JHEP} {\bfseries 02} (2023)
  142}, [\href{http://arxiv.org/abs/2211.01729}{{\ttfamily arXiv:2211.01729}}
  [hep-ph]]. [Addendum: JHEP 06, 119 (2024)].

\bibitem{Esteban:2021ozz}
Ivan Esteban and Jordi Salvado, ``{\it {Long Range Interactions in Cosmology:
  Implications for Neutrinos}},''
  \href{http://dx.doi.org/10.1088/1475-7516/2021/05/036}{{JCAP} {\bfseries 05}
  (2021) 036}, [\href{http://arxiv.org/abs/2101.05804}{{\ttfamily
  arXiv:2101.05804}} [hep-ph]].

\bibitem{Ge:2023nnh}
Shao-Feng Ge, Pedro Pasquini, and Liang Tan, ``{\it {Neutrino mass measurement
  with cosmic gravitational focusing}},''
  \href{http://dx.doi.org/10.1088/1475-7516/2024/05/108}{{JCAP} {\bfseries 05}
  (2024) 108}, [\href{http://arxiv.org/abs/2312.16972}{{\ttfamily
  arXiv:2312.16972}} [hep-ph]].

\bibitem{Zhu:2013tma}
Hong-Ming Zhu, Ue-Li Pen, Xuelei Chen, Derek Inman, and Yu~Yu, ``{\it
  {Measurement of Neutrino Masses from Relative Velocities}},''
  \href{http://dx.doi.org/10.1103/PhysRevLett.113.131301}{{Phys. Rev. Lett.}
  {\bfseries 113} (2014) 131301},
  [\href{http://arxiv.org/abs/1311.3422}{{\ttfamily arXiv:1311.3422}}
  [astro-ph.CO]].

\bibitem{Zhu:2014qma}
Hong-Ming Zhu, Ue-Li Pen, Xuelei Chen, and Derek Inman, ``{\it {Probing
  Neutrino Hierarchy and Chirality via Wakes}},''
  \href{http://dx.doi.org/10.1103/PhysRevLett.116.141301}{{Phys. Rev. Lett.}
  {\bfseries 116} no.~14, (2016) 141301},
  [\href{http://arxiv.org/abs/1412.1660}{{\ttfamily arXiv:1412.1660}}
  [astro-ph.CO]].

\bibitem{Inman:2015pfa}
Derek Inman, J.~D. Emberson, Ue-Li Pen, Alban Farchi, Hao-Ran Yu, and Joachim
  Harnois-D\'eraps, ``{\it {Precision reconstruction of the cold dark
  matter-neutrino relative velocity from $N$-body simulations}},''
  \href{http://dx.doi.org/10.1103/PhysRevD.92.023502}{{Phys. Rev. D} {\bfseries
  92} no.~2, (2015) 023502}, [\href{http://arxiv.org/abs/1503.07480}{{\ttfamily
  arXiv:1503.07480}} [astro-ph.CO]].

\bibitem{Okoli:2016vmd}
Chiamaka Okoli, Morag~I. Scrimgeour, Niayesh Afshordi, and Michael~J. Hudson,
  ``{\it {Dynamical friction in the primordial neutrino sea}},''
  \href{http://dx.doi.org/10.1093/mnras/stx560}{{Mon. Not. Roy. Astron. Soc.}
  {\bfseries 468} no.~2, (2017) 2164--2175},
  [\href{http://arxiv.org/abs/1611.04589}{{\ttfamily arXiv:1611.04589}}
  [astro-ph.CO]].

\bibitem{Zhu:2019kzb}
Hong-Ming Zhu and Emanuele Castorina, ``{\it {Measuring dark matter-neutrino
  relative velocity on cosmological scales}},''
  \href{http://dx.doi.org/10.1103/PhysRevD.101.023525}{{Phys. Rev. D}
  {\bfseries 101} no.~2, (2020) 023525},
  [\href{http://arxiv.org/abs/1905.00361}{{\ttfamily arXiv:1905.00361}}
  [astro-ph.CO]].

\bibitem{Nascimento:2023ezc}
Caio Nascimento and Marilena Loverde, ``{\it {Neutrino winds on the sky}},''
  \href{http://dx.doi.org/10.1088/1475-7516/2023/11/036}{{JCAP} {\bfseries 11}
  (2023) 036}, [\href{http://arxiv.org/abs/2307.00049}{{\ttfamily
  arXiv:2307.00049}} [astro-ph.CO]].

\bibitem{Ge:2019ldu}
Shao-Feng Ge and Jing-Yu Zhu, ``{\it {Phenomenological Advantages of the Normal
  Neutrino Mass Ordering}},''
  \href{http://dx.doi.org/10.1088/1674-1137/44/8/083103}{{Chin. Phys. C}
  {\bfseries 44} no.~8, (2020) 083103},
  [\href{http://arxiv.org/abs/1910.02666}{{\ttfamily arXiv:1910.02666}}
  [hep-ph]].

\bibitem{Planck:2019nip}
{\bfseries Planck} Collaboration, N.~Aghanim {et~al.}, ``{\it {Planck 2018
  results. V. CMB power spectra and likelihoods}},''
  \href{http://dx.doi.org/10.1051/0004-6361/201936386}{{Astron. Astrophys.}
  {\bfseries 641} (2020) A5},
  [\href{http://arxiv.org/abs/1907.12875}{{\ttfamily arXiv:1907.12875}}
  [astro-ph.CO]].

\bibitem{Carron:2022eyg}
Julien Carron, Mark Mirmelstein, and Antony Lewis, ``{\it {CMB lensing from
  Planck PR4~maps}},''
  \href{http://dx.doi.org/10.1088/1475-7516/2022/09/039}{{JCAP} {\bfseries 09}
  (2022) 039}, [\href{http://arxiv.org/abs/2206.07773}{{\ttfamily
  arXiv:2206.07773}} [astro-ph.CO]].

\bibitem{ACT:2023kun}
{\bfseries ACT} Collaboration, Mathew~S. Madhavacheril {et~al.}, ``{\it {The
  Atacama Cosmology Telescope: DR6 Gravitational Lensing Map and Cosmological
  Parameters}},'' \href{http://dx.doi.org/10.3847/1538-4357/acff5f}{{Astrophys.
  J.} {\bfseries 962} no.~2, (2024) 113},
  [\href{http://arxiv.org/abs/2304.05203}{{\ttfamily arXiv:2304.05203}}
  [astro-ph.CO]].

\bibitem{ACT:2023dou}
{\bfseries ACT} Collaboration, Frank~J. Qu {et~al.}, ``{\it {The Atacama
  Cosmology Telescope: A Measurement of the DR6 CMB Lensing Power Spectrum and
  Its Implications for Structure Growth}},''
  \href{http://dx.doi.org/10.3847/1538-4357/acfe06}{{Astrophys. J.} {\bfseries
  962} no.~2, (2024) 112}, [\href{http://arxiv.org/abs/2304.05202}{{\ttfamily
  arXiv:2304.05202}} [astro-ph.CO]].

\bibitem{ACT:2023ubw}
{\bfseries ACT} Collaboration, Niall MacCrann {et~al.}, ``{\it {The Atacama
  Cosmology Telescope: Mitigating the Impact of Extragalactic Foregrounds for
  the DR6 Cosmic Microwave Background Lensing Analysis}},''
  \href{http://dx.doi.org/10.3847/1538-4357/ad2610}{{Astrophys. J.} {\bfseries
  966} no.~1, (2024) 138}, [\href{http://arxiv.org/abs/2304.05196}{{\ttfamily
  arXiv:2304.05196}} [astro-ph.CO]].

\bibitem{Ravoux:2024gtk}
{\bfseries DESI Lyman-alpha working group} Collaboration, Corentin Ravoux,
  ``{\it {One-dimensional power spectrum from first DESI
  Lyman-\ensuremath{\alpha} forest}},'' in {{58th Rencontres de Moriond on
  Cosmology}}.
\newblock 5, 2024.
\newblock [\href{http://arxiv.org/abs/2405.03447}{{\ttfamily arXiv:2405.03447}}
  [astro-ph.CO]].

\bibitem{KamLAND:2013rgu}
{\bfseries KamLAND} Collaboration, A.~Gando {et~al.}, ``{\it {Reactor On-Off
  Antineutrino Measurement with KamLAND}},''
  \href{http://dx.doi.org/10.1103/PhysRevD.88.033001}{{Phys. Rev. D} {\bfseries
  88} no.~3, (2013) 033001}, [\href{http://arxiv.org/abs/1303.4667}{{\ttfamily
  arXiv:1303.4667}} [hep-ex]].

\bibitem{T2K:2020nqo}
{\bfseries T2K} Collaboration, K.~Abe {et~al.}, ``{\it {T2K measurements of
  muon neutrino and antineutrino disappearance using $3.13\times 10^{21}$
  protons on target}},''
  \href{http://dx.doi.org/10.1103/PhysRevD.103.L011101}{{Phys. Rev. D}
  {\bfseries 103} no.~1, (2021) L011101},
  [\href{http://arxiv.org/abs/2008.07921}{{\ttfamily arXiv:2008.07921}}
  [hep-ex]].

\bibitem{LoVerde:2014pxa}
Marilena LoVerde, ``{\it {Halo bias in mixed dark matter cosmologies}},''
  \href{http://dx.doi.org/10.1103/PhysRevD.90.083530}{{Phys. Rev. D} {\bfseries
  90} no.~8, (2014) 083530}, [\href{http://arxiv.org/abs/1405.4855}{{\ttfamily
  arXiv:1405.4855}} [astro-ph.CO]].

\bibitem{Ge:2016tfx}
Shao-Feng Ge and Manfred Lindner, ``{\it {Extracting Majorana properties from
  strong bounds on neutrinoless double beta decay}},''
  \href{http://dx.doi.org/10.1103/PhysRevD.95.033003}{{Phys. Rev. D} {\bfseries
  95} no.~3, (2017) 033003}, [\href{http://arxiv.org/abs/1608.01618}{{\ttfamily
  arXiv:1608.01618}} [hep-ph]].

\bibitem{Agostini:2022zub}
Matteo Agostini, Giovanni Benato, Jason~A. Detwiler, Javier Men\'endez, and
  Francesco Vissani, ``{\it {Toward the discovery of matter creation with
  neutrinoless \ensuremath{\beta}\ensuremath{\beta} decay}},''
  \href{http://dx.doi.org/10.1103/RevModPhys.95.025002}{{Rev. Mod. Phys.}
  {\bfseries 95} no.~2, (2023) 025002},
  [\href{http://arxiv.org/abs/2202.01787}{{\ttfamily arXiv:2202.01787}}
  [hep-ex]].

\bibitem{Mei:2024kvs}
Dongming Mei, Kunming Dong, Austin Warren, and Sanjay Bhattarai, ``{\it {Impact
  of recent updates to neutrino oscillation parameters on the effective
  Majorana neutrino mass in
  0\ensuremath{\nu}\ensuremath{\beta}\ensuremath{\beta} decay}},''
  \href{http://dx.doi.org/10.1103/PhysRevD.110.015026}{{Phys. Rev. D}
  {\bfseries 110} no.~1, (2024) 015026},
  [\href{http://arxiv.org/abs/2404.19624}{{\ttfamily arXiv:2404.19624}}
  [hep-ph]].

\bibitem{Cao:2019hli}
Jun Cao, Guo-Yuan Huang, Yu-Feng Li, Yifang Wang, Liang-Jian Wen, Zhi-Zhong
  Xing, Zhen-Hua Zhao, and Shun Zhou, ``{\it {Towards the meV limit of the
  effective neutrino mass in neutrinoless double-beta decays}},''
  \href{http://dx.doi.org/10.1088/1674-1137/44/3/031001}{{Chin. Phys. C}
  {\bfseries 44} no.~3, (2020) 031001},
  [\href{http://arxiv.org/abs/1908.08355}{{\ttfamily arXiv:1908.08355}}
  [hep-ph]].

\bibitem{DESI:2025ejh}
{\bfseries DESI} Collaboration, W.~Elbers {et~al.}, ``{\it {Constraints on
  Neutrino Physics from DESI DR2 BAO and DR1 Full Shape}},''
  [\href{http://arxiv.org/abs/2503.14744}{{\ttfamily arXiv:2503.14744}}
  [astro-ph.CO]].

\end{thebibliography}\endgroup

\end{document}